\documentclass{ws-procs961x669}            % default, citations in superscript

\def\Lie{\pounds}
\def\dual{{}^\star\!}

\begin{document}
\title{Covariant Hamiltonian boundary term: \\ Reference and quasi-local quantities}

\author{Gang Sun$^1$, Chiang-Mei Chen$^2$, Jian-Liang Liu$^3$ and James M. Nester$^{2,4,5}$}
\address{
$^1$Institute of Theoretical Physics, Beijing University of Technology, Beijing, 100136, China \\
$^2$Department of Physics, National Central University, Chungli 32001, Taiwan \\
%$^?$Center for Mathematical and Theoretical Physics, National Central University, Chungli 32001, Taiwan \\
$^3$Department of Mathematics, Shantou University, Guangdong 515063, China \\
$^4$Graduate Institute of Astronomy, National Central University, Chungli 32001, Taiwan \\
$^5$Leung Center for Cosmology and Particle Astrophysics, National Taiwan University, \\ Taipei 10617, Taiwan \\
sungang@bjut.edu.cn, cmchen@phy.ncu.edu.tw, liujl\_mc@163.com, nester@phy.ncu.edu.tw}

%\author{Jian-Liang Liu}
%\address{$^2$ Department of Mathematics, Shantou University, Guangdong 515063, China \\
%E-mail: liujl\_mc@163.com }

%\author{James M. Nester}
%\address{Department of Physics, National Central University, Chungli 32001, Taiwan \\
%$^3$ Leung Center for Cosmology and Particle Astrophysics, National Taiwan University, Taipei 10617, Taiwan \\
%E-mail: nester@phy.ncu.edu.tw}

%\author{Gang Sun}
%\address{$^4$ Institute of Theoretical Physics, Beijing University of Technology, Beijing, 100136, China \\
%E-mail: sungang@bjut.edu.cn}

\begin{abstract}
The Hamiltonian for dynamic geometry generates the evolution of a spatial region along a vector field. It includes a boundary term which determines both the value of the Hamiltonian and the boundary conditions. The value gives the quasi-local quantities: energy-momentum, angular-momentum and center-of-mass. The boundary term depends not only on the dynamical variables but also on their reference values; the latter determine the ground state (having vanishing quasi-local quantities). For our preferred boundary term for Einstein's GR we propose 4D isometric matching and extremizing the energy to determine the reference metric and connection values.
\end{abstract}


\bodymatter

\section{Introduction}
Although the global \emph{total} energy-momentum is well defined (for spaces with suitable asymptotic regions), for any gravitating system --- and hence for all real \emph{physical} systems --- the localization of energy-momentum is still an outstanding fundamental problem.~\cite{Sza09, Chen:2015vya} Unlike all matter and other interaction fields, the gravitational field itself has \emph{no} proper energy-momentum density. In view of the fact that energy-momentum is conserved, and that sources \emph{exchange} energy-momentum \emph{locally} with the gravitational field, one expects some kind of ``local description'' of the energy-momentum density of gravity itself. But all attempts at constructing such an expression led only to reference frame dependent quantities, generally referred to as \emph{pseudotensors}.~\cite{Chang:1998wj} Physically this can be understood as a consequence of \emph{Einstein's equivalence principle}: gravity cannot be detected at a point. The energy-momentum of gravity --- and thus for all physical systems is inherently \emph{non-local}. The modern idea is \emph{quasi-local}: energy-momentum is associated with a closed surface bounding a region.~\cite{Penrose82}
%\textbf{We will focus on Chen-Nester-Tung (CNT) quasi-local formalism in the following.}

\section{Covariant Hamiltonian Formalism}
%\subsection{First order Lagrangian}
The \emph{first order} Lagrangian~\cite{Kuchar} for an \emph{$f$-form field $\varphi$} and its \emph{conjugate momentum $p$} is given by
\begin{equation}
{\cal L} = \mathcal{L}(d\varphi; \varphi, p) = d \varphi \wedge p - \Lambda(\varphi, p). \label{1st Lagrangian}
\end{equation}
The variation (with respect to $\varphi$ and $p$ independently)
\begin{equation}
\delta {\cal L} = d(\delta \varphi \wedge p) + \delta \varphi \wedge \frac{\delta {\cal L}}{\delta \varphi} + \frac{\delta {\cal L}}{\delta p} \wedge \delta p
\end{equation}
gives the equations of motion, with $\varsigma := (-1)^f$,
\begin{equation}
\frac{\delta {\cal L}}{\delta p} := d \varphi - \partial_p \Lambda = 0, \qquad \frac{\delta {\cal L}}{\delta \varphi} := - \varsigma d p - \partial_\varphi \Lambda = 0.
\end{equation}
%where $\varsigma := (-1)^f$.

Diffeomorphism invariance (in terms of the Lie derivative $\Lie_N = d i_N + i_N d$) leads an identity for \emph{any vector $N$}
\begin{equation}
d i_N {\cal L} \equiv \Lie_N {\cal L} \equiv d( \Lie_N \varphi \wedge p) + \Lie_N \varphi \wedge \frac{\delta {\cal L}}{\delta \varphi} + \frac{\delta {\cal L}}{\delta p} \wedge \Lie_N p.
\end{equation}
From this one gets a conserved ``translational current'' 3-form:
\begin{equation}
{\cal H}(N) := \Lie_N \varphi \wedge p - i_N {\cal L}, \qquad - d {\cal H}(N) \equiv \Lie_N \varphi \wedge \frac{\delta {\cal L}}{\delta \varphi} + \frac{\delta {\cal L}}{\delta p} \wedge \Lie_N p.
\end{equation}
Note that ${\cal H}(N)$ is not unique:
\begin{equation}
\mathcal{H}' = \mathcal{H} + d \mathcal{B}' \quad \Rightarrow \quad d \mathcal{H} = d \mathcal{H}'.
\end{equation}
Furthermore it can be written in the form, ${\cal H}(N) = N^\mu {\cal H}_\mu + d {\cal B}(N)$ then
\begin{eqnarray}
&& d{\cal H}(N) = d [ N^\mu {\cal H}_\mu + d {\cal B}(N) ] \equiv d N^\mu \wedge {\cal H}_\mu + N^\mu d {\cal H}_\mu
\\
\Rightarrow && {\cal H}_\mu \; \textrm{vanishes ``on shell''}. \nonumber
\end{eqnarray}
Hence for gravitating systems the Noether translational ``charge'' --- \emph{energy-momentum} --- is \emph{quasi-local}, it is given by the integral of the boundary term, ${\cal B}(N)$. But this boundary term can be completely modified to any value. However, the Hamiltonian approach tames the ambiguity. Quasi-local quantities are determined only by the surface integral
\begin{equation}
E(N) = \int_\Sigma {\cal H}(N) = \int_\Sigma \left[ N^\mu {\cal H}_\mu + d {\cal B}(N) \right] = \oint_{\partial\Sigma} {\cal B}(N).
\end{equation}
The two parts of the Hamiltonian have distinct roles: The 3-form part $N^\mu \mathcal{H}_\mu$ generates the equations of motion. As mentioned, for diffeomorphic invariant theories it has vanishing value. The Hamiltonian generally also includes a boundary term ${\cal B}(N)$: (i) it determines the values of the quasi-local quantities, and (ii) it determines the boundary conditions.~\cite{Chen:1994qg, Chen:1998aw}

\subsection{Quasi-local Quantities}
The Hamiltonian boundary terms determines the values of the quasi-local quantities:
\begin{itemize}
  \item Energy is given by a suitable \emph{timelike} displacement;

  \item Linear momentum is obtained from a \emph{spatial} translation;

  \item Angular momentum from a suitable \emph{rotational} displacement;

  \item A spacetime displacement which is asymptotically a \emph{boost} will give the  center-of-mass moment.
\end{itemize}
Our Noether analysis has revealed that ${\cal B}(N)$ can be adjusted, changing the conserved value to a new value. However the variational principle contains an additional (largely overlooked) feature which distinguishes all of these choices. The boundary variation principle, i.e.\ the boundary term in the variation, tells us what to hold fixed on the boundary --- it determines the \emph{boundary conditions}.~\cite{Chang:1998wj}

The different Hamiltonian boundary terms are each associated with distinct boundary conditions. As in thermodynamics or electrostatics there are various ``energies'' which correspond to how the system interacts with the outside through its boundary. In general (in particular for gravity) it is necessary (in order to guarantee functional differentiability of the Hamiltonian on the phase space with the desired boundary conditions) to adjust the boundary term ${\cal B}(N) = i_N \varphi \wedge p$ which is naturally inherited from the Lagrangian~\eqref{1st Lagrangian}. The variation of the Hamiltonian implies
\begin{equation}
\delta {\cal H}(N) \equiv - \delta \varphi \wedge \Lie_N p + \Lie_N \varphi \wedge \delta p + d i_N(\delta \varphi \wedge p) - i_N \left( \delta \varphi \wedge \frac{\delta {\cal L}}{\delta \varphi} + \frac{\delta {\cal L}}{\delta p} \wedge \delta p \right).
\end{equation}
There is a freedom for modifying the boundary term ${\cal B}(N) \to {\cal B}'(N)$. Moreover, a reference configuration, $\bar\varphi$ and $\bar p$, (which determines the ground state) is essential especially for gravity, in particular to allow the desired phase space asymptotics.

With $\Delta \varphi := \varphi - \bar\varphi$, $\Delta p := p - \bar p$, we found two boundary choices (essentially \emph{Dirichlet} and \emph{Neumann}) which have the indicated covariant boundary terms in $\delta{\cal H}$:
\begin{eqnarray}
{\cal B}_{\varphi} &=& i_N \varphi \wedge \Delta p - \varsigma \Delta \varphi \wedge i_N \bar p \quad \Longrightarrow \quad i_N (\delta \varphi \wedge \Delta p),
\\
{\cal B}_{p} &=& i_N \bar \varphi \wedge \Delta p - \varsigma \Delta \varphi \wedge i_N p \quad \Longrightarrow \; - i_N (\Delta \varphi \wedge \delta p).
\end{eqnarray}
We also found two other physical interesting choices:
\begin{eqnarray}
{\cal B}_{\mathrm{dynamics}} &=& i_N \bar\varphi \wedge \Delta p - \varsigma \Delta \varphi \wedge i_N \bar p \quad \Longrightarrow \quad \varsigma \delta \varphi \wedge i_N \Delta p - i_N \Delta \varphi \wedge \delta p,
\\
{\cal B}_{\mathrm{constraint}} &=& i_N \varphi \wedge \Delta p - \varsigma \Delta \varphi \wedge i_N p \quad \Longrightarrow \quad i_N \delta \varphi \wedge \Delta p - \varsigma \Delta \varphi \wedge i_N \delta p.
\end{eqnarray}
Let us look at the following two applications.

\subsection{Applications: Electromagnetism and General Relativity}
The first order Lagrangian 4-form for the source free $U(1)$ gauge field one-form $A$ and its conjugate momentum $H$ is
\begin{equation}
{\cal L}_{\mathrm{EM}} = d A \wedge H - \frac{1}{2} \dual H \wedge H.
\end{equation}
The pair of first order equations are
\begin{equation}
d H = 0, \qquad d A - \dual H = 0.
\end{equation}
These are just the vacuum Maxwell equations with $\dual H = F:= dA$; hence $H = - \dual F$ and $d \dual F = 0$. The natural reference in electromagnetism is $\bar A = 0, \; \bar H = 0$. The best boundary choice is ${\cal B}_{\rm dynamics}$, which vanishes for this reference choice.

%\subsection{Application to General Relativity}
The first order Lagrangian for Einstein's (vacuum) gravity theory is
\begin{equation}
{\cal L}_{\rm GR} = R^\alpha{}_\beta \wedge \eta_\alpha{}^\beta,
\end{equation}
where the curvature 2-form is $R^\alpha{}_\beta := d \Gamma^\alpha{}_\beta + \Gamma^\alpha{}_\gamma \wedge \Gamma^\gamma{}_\beta$, and the dual basis 2-form is $\eta^{\alpha\beta} := *(\vartheta^\alpha \wedge \vartheta^\beta)$. For gravity itself, two different choices of boundary condition correspond to the quasi-local expressions which asymptotically give (a) the ADM energy, (b) the Bondi energy and, moreover, (c) the Bondi flux: the celebrated outgoing flux plus an incoming flux.~\cite{Chen:2005hwa} Our general formalism with $\varphi \to \Gamma^\alpha{}_\beta$ and $p \to \eta_\alpha{}^\beta$ gives 4 quasi-local expressions.
%Each has its associated energy flux formula.
There is a distinguished energy expression with a very desirable property: it corresponds to imposing boundary conditions on a \emph{manifestly covariant object}:
\begin{equation}
{\cal B}_\vartheta({\bf N}) := \frac1{16 \pi} \left( \Delta \Gamma^\alpha{}_\beta \wedge i_N \eta_\alpha{}^\beta + {\bar D}^\alpha N^\beta \Delta \eta_{\alpha\beta} \right).
\end{equation}
The associated energy flux expression is
\begin{equation}
\Lie_N {\cal H}_{\vartheta} \simeq d i_N \left( \Delta \Gamma^{\alpha\beta} \wedge \Lie_N \eta_{\alpha\beta} \right),
\end{equation}
and the natural reference in gravity for the asymptotic flat spacetime is the Minkowski spacetime:
\begin{equation}
\bar g_{\mu\nu} = \mathrm{diag} (-1, +1, +1, +1).
\end{equation}

\section{Reference Choice}
Now let us turn to how to select the reference; effectively one should embed the 2-boundary into Minkowski space.~\cite{Wu:2012mi, Liu:2011jha}
%Now let's turn to the embedding problem and see how to decide the reference geometry~\cite{Wu:2012mi, Liu:2011jha}.
In a neighborhood of the desired spacelike boundary 2-surface $S$, 4 smooth functions $y^i = y^i(x^\mu), \; i = 0, 1, 2, 3$ with $dy^0 \wedge dy^1 \wedge dy^2 \wedge dy^3 \ne 0$ define a Minkowski reference:
\begin{equation}
\bar g = -(dy^0)^2 + (dy^1)^2 + (dy^2)^2 + (dy^3)^2.
\end{equation}
The reference connection is
\begin{equation}
\bar \Gamma^\alpha{}_\beta = x^\alpha{}_i ( \bar\Gamma^i{}_j y^j{}_\beta + dy^i{}_\beta ) = x^\alpha{}_i dy^i{}_\beta,
\end{equation}
where $dy^i = y^i{}_\alpha dx^\alpha$ and $dx^\alpha = x^\alpha{}_j dy^j$ with vanishing Minkowski reference connection coefficients. $N^\mu$ is a \emph{translational Killing field} of the Minkowski reference, then the second quasi-local term vanishes. Our quasi-local expression then takes the form
\begin{equation}
{\mathcal B}(N) = N^k x^\mu{}_k (\Gamma^\alpha{}_\beta - x^\alpha{}_j \, dy^j{}_\beta) \wedge \eta_{\mu\alpha}{}^\beta.
\end{equation}
To determine the reference choice $y^{i}{}_{\mu}$ in terms of quasi-spherical foliation adapted coordinates $t, r, \theta, \phi$, the isometric matching on the 2-surface implies
\begin{equation}
g_{AB} = \bar g_{AB} = \bar g_{ij} y^i_A y^j_B = - y^0_A y^0_B + \delta_{ab} y^a_A y^b_B, \quad a, b = 1, 2, 3; \; A, B = 2, 3 = \theta, \phi,
\end{equation}
where the reference metric on the dynamical space has the components $\bar g_{\mu\nu} = \bar g_{ij} y^i{}_\mu y^j{}_\nu$. From a classic closed 2-surface into $\mathbb R^3$ embedding theorem, we expect that---as long as one restricts $S$ and $y^0(x^A)$ such that on $S$
\begin{equation}
g_{AB}' := g_{AB} + y^0_A y^0_B
\end{equation}
is convex---one can prove that there is a unique isometric embedding. (But, unfortunately, there is no explicit formula.)

\subsection{4D Isometric Matching}
Complete 4D isometric matching on $S$ has 10 constraints:~\cite{Liu:2011jha}
\begin{equation}
g_{\mu\nu}|_S = \bar g_{\mu\nu}|_S = \bar g_{ij} y^i{}_\mu y^j{}_\nu|_S.
\end{equation}
There are 12 embedding functions on the constant $t, r$ 2-surface:
\begin{equation}
y^i (\Rightarrow y^i_\theta, y^i_\phi), \quad y^i_t, \quad y^i_r.
\end{equation}
The 10 constraints split into 3 for the already discussed 2D isometric matching: $g_{\theta\theta}, g_{\theta\varphi}, g_{\varphi\varphi}$ which constrain the 4 $y^i$; 3 normal bundle algebraic quadratic expressions: $g_{tt}, g_{tr}, g_{rr}$; and 4 mixed linear algebraic expressions: $g_{t\theta}, g_{t\varphi}, g_{r\theta}, g_{r\varphi}$. The 2D isometric matching can be regarded as a given $y^0$ uniquely determining $y^1,y^2,y^3$ on $S$. The remaining 7 algebraic equations can be regarded as finding all the other embedding variables in terms of $y^i$ and $y^0{}_r$ on $S$. Thus one can take $y^0,y^0{}_r$ as the embedding control variables. Geometrically $y^0{}_r$ controls a boost in the plane normal to $S$.

\subsection{An Optimal Choice}
One can regard the value of the boundary term as a measure of the difference between the dynamical boundary values and the reference boundary values. However, how to find the ``best matched'' reference geometry? Because 12 embedding variables are subject to 10 isometric conditions, one will obtain the best matched reference geometry as long as one can obtain the two unknown variables.~\cite{Chen:2009zd, Wu:2011wk, Sun:2013ika}

For a given $S$ there are 2 different quantities which can be considered: $m^2 = - \bar g{}^{ij} p_i p_j$ and $E(N, S)$. For the latter there are 2 different ways to fix $N$. The critical points are distinguished: (1) The critical points of $m^2$. This determines the reference up to Poincar\'e transformations. (2) The critical points of $E(\partial_T, S)$. (If $m^2 > 0$, this may equivalent to (1).) (3) The critical points of $E(N, S)$ for a given dynamical vector field $N$. (Afterward one could find the extreme choice of $N$.)

The first two approaches lead to quasi-local quantities associated with $S$, the third alternative gives a quasi-local energy associated with an observer.~\cite{Nester:2012zi, Sun:2015mxg} Based on some physical and practical computational arguments, it is reasonable to expect a unique solution. %[Numerically compute many cases, the critical values stand out.]
For our quasi-local values for axisymmetric solutions including Kerr see Ref.~\refcite{Sun:2013ika}.

\section{Summary}
For any gravitating system --- and hence for all \emph{physical} systems --- the \emph{localization} of energy-momentum is an outstanding problem. We've displayed the relation between covariant Hamiltonian boundary term and the quasi-local quantities. %by CNT formalism.
For gravitating systems, we have obtained four quasi-local energy-momentum expressions; each is associated with a physically distinct, and geometrically clear, boundary condition. With the ``\emph{best matched}'' reference, we have a satisfactory way of fixing the Hamiltonian boundary term quasi-locally for locally Poincar\'e gauge invariant gravity including GR. This in particular gives a way of resolving the ambiguities in determining the quasi-local energy-momentum of classical physical systems.

\section*{Acknowledgement}
C.M.C. was supported by the Ministry of Science and Technology of the R.O.C. under the grant MOST 102-2112-M-008-015-MY3.


\begin{thebibliography}{10}

\bibitem{Sza09}
  L.~B.~Szabados,
%  ``Quasi-Local Energy-Momentum and Angular Momentum in General Relativity,''
  \emph{Living Rev.\ Relativ.} {\bf 12}, 4 (2009). %http://www.livingreviews.org/lrr-2009-4.

\bibitem{Chen:2015vya}
  C.-M.~Chen, J.~M.~Nester and R.-S.~Tung,
%  ``Gravitational energy for GR and Poincare gauge theories: A covariant Hamiltonian approach,''
  \emph{Int.\ J.\ Mod.\ Phys.\ D} {\bf 24}, 1530026 (2015) [arXiv:1507.07300 [gr-qc]].

\bibitem{Chang:1998wj}
  C.-C.~Chang, J.~M.~Nester and C.-M.~Chen,
%  ``Pseudotensors and Quasilocal Gravitational Energy Momentum,''
  \emph{Phys.\ Rev.\ Lett.} {\bf 83}, 1897--1901 (1999) [gr-qc/9809040].

\bibitem{Penrose82}
  R.~Penrose,
%  ``Quasi-local Mass and Angular Momentum in General Relativity,''
  \emph{Proc.\ R.\ Soc.\ London} \ A {\bf 381}, 53--63 (1982).

\bibitem{Kuchar}
  K.~Kucha\v{r},
%  ``Dynamics of Tensor Fields in Hyperspace. III,''
  \emph{J.\ Math.\ Phys.} {\bf 17} 801--820 (1976).

\bibitem{Chen:1994qg}
  C.-M.~Chen, J.~M.~Nester and R.-S.~Tung,
%  ``Quasilocal Energy Momentum for Gravity Theories,''
 \emph{Phys.\ Lett.\ A} {\bf 203}, 5--11 (1995) [gr-qc/9411048].

\bibitem{Chen:1998aw}
  C.-M.~Chen and J.~M.~Nester,
%  ``Quasilocal Quantities for GR and other Gravity Theories,''
  \emph{Class.\ Quantum Grav.} {\bf 16}, 1279--1304 (1999) [gr-qc/9809020].

\bibitem{Chen:2005hwa}
  C.-M.~Chen, J.~M.~Nester and R.-S.~Tung,
%  ``The Hamiltonian Boundary Term and Quasi-local Energy Flux,''
  \emph{Phys.\ Rev.\ D} {\bf 72}, 104020 (2005) [gr-qc/0508026].

\bibitem{Liu:2011jha}
  J.-L.~Liu, C.-M.~Chen and J.-M.~Nester,
%  ``Quasi-local energy and the choice of reference,''
  \emph{Class.\ Quant.\ Grav.} {\bf 28}, 195019 (2011) [arXiv:1105.0502 [gr-qc]].

\bibitem{Wu:2012mi}
  M.-F.~Wu, C.-M.~Chen, J.-L.~Liu and J.~M.~Nester,
%  ``Quasi-local Energy for Spherically Symmetric Spacetimes,''
  \emph{Gen.\ Rel.\ Grav.} {\bf 44}, 2401 (2012) [arXiv:1206.0506 [gr-qc]].

\bibitem{Chen:2009zd}
  C.-M.~Chen, J.-L.~Liu, J.~M.~Nester and M.-F.~Wu,
%  ``Optimal Choices of Reference for Quasi-local Energy,''
  \emph{Phys.\ Lett.\ A} {\bf 374}, 3599 (2010) [arXiv:0909.2754 [gr-qc]].

\bibitem{Wu:2011wk}
  M.-F.~Wu, C.-M.~Chen, J.-L.~Liu and J.~M.~Nester,
%  ``Optimal Choices of Reference for a Quasi-local Energy: Spherically Symmetric Spacetimes,''
  \emph{Phys.\ Rev.\ D} {\bf 84}, 084047 (2011) [arXiv:1109.4738 [gr-qc]].

\bibitem{Sun:2013ika}
  G.~Sun, C.-M.~Chen, J.-L.~Liu and J.~M.~Nester,
%  ``An Optimal Choice of Reference for the Quasi-Local Gravitational Energy and Angular Momentum,''
  \emph{Chin.\ J.\ Phys.} {\bf 52}, 111 (2014) [arXiv:1307.1039 [gr-qc]].

\bibitem{Nester:2012zi}
  J.~M.~Nester, C.-M.~Chen, J.-L.~Liu and G.~Sun,
%  ``A reference for the covariant Hamiltonian boundary term,''
  Springer Proc.\ Phys.\  {\bf 157}, 177 (2014) [arXiv:1210.6148 [gr-qc]].

\bibitem{Sun:2015mxg}
  G.~Sun, C.-M.~Chen, J.-L.~Liu and J.~M.~Nester,
%  ``A Reference for the Gravitational Hamiltonian Boundary Term,''
  \emph{Chin.\ J.\ Phys.} {\bf 53}, 110107 (2015).

\end{thebibliography}
\end{document}